\documentclass{emulateapj}
\usepackage{natbib}

\def\gsim{\;\rlap{\lower 2.5pt
 \hbox{$\sim$}}\raise 1.5pt\hbox{$>$}\;}
\def\lsim{\;\rlap{\lower 2.5pt
   \hbox{$\sim$}}\raise 1.5pt\hbox{$<$}\;}
\def\msolarunits{$h^{-1}$ $M_{\odot}$ }
\def\lengthunits{$h^{-1}$ kpc }
\def\velocityunits{km s$^{-1}$ }
\def\timeunits{$h^{-1}$ Gyr }

\begin{document}

\title{Antitruncated Stellar Disks via Minor Mergers}
\author{Joshua D. Younger, T. J. Cox, Anil C. Seth, \& Lars Hernquist}

\affil{Harvard--Smithsonian Center for Astrophysics,
	 60 Garden Street, Cambridge, MA 02138}

\email{jyounger@cfa.harvard.edu}

\begin{abstract}

We use hydrodynamic simulations of minor mergers of galaxies to
investigate the nature of surface brightness excesses at large radii
observed in some spiral galaxies: antitruncated stellar disks.  We find that
this process can produce the antitruncation via two competing
effects: (1) merger-driven gas inflows that concentrate mass in the
center of the primary galaxy and contract its inner density profile;
and (2) angular momentum transferred outwards by the interaction,
causing the outer disk to expand.  In our experiments, this requires
both a significant supply of gas in the primary disk, and that the
encounter be prograde with moderate orbital angular momentum.  The
stellar surface mass density profiles of our remnants both
qualitatively and quantitatively resemble the broken exponentials
observed in local face--on spirals that display antitruncations.
Moreover, the observed trend towards more frequent antitruncation
relative to classical truncation in earlier Hubble types is consistent
with a merger-driven scenario.

\end{abstract}

\keywords{galaxies: kinematics and dynamics, galaxies: interactions,
galaxies: formation, galaxies: evolution, galaxies: spiral,
galaxies: structure, methods: $n$--body simulations}

\section{Introduction}

Through the pioneering studies of \citet{patterson1940} and
\citet{deVaucouleurs1959}, it was first recognized that the stellar
disks of most spiral galaxies are well-approximated by an exponential
surface brightness profile.  However, \citet{vanderKruit1979} later
found that in many cases this description breaks down at large radius,
where the disk surface density appears to be truncated
\citep[see][and references therein]{pohlen2006}.

Many studies of truncated disks have examined the surface brightness
profiles of local edge--on systems \citep{vanderKruit1979,barnaby1992,
barteldrees1994, pohlen2000,deGrijs2001,vanderKruit2001,kregel2002,Florido2006b,Florido2006a} and at higher redshifts $z\sim 1$ \citep{perez2004,trujillo2005}.  This
choice of inclination facilitates detection of truncations, but is
subject to potential biases owing to the effects of dust extinction
and line--of--sight integration \citep[see, e.g.,][]{pohlen2006}.
Observations of face--on systems, which mitigate these complications,
were also successful at detecting truncation, but found that the
surface brightness profiles are better represented by a broken
exponential than a hard break \citep{pohlen2002,erwin2005,pohlen2006}.  \citet{hunter2006}
also note the existence of double--exponentials in their sample of 
very late--type spirals and dwarfs.

These studies also uncovered a broad range of behaviors, including
some disks that follow a pure exponential profile out to very large
radius \citep{BlandHawthorn2005,pohlen2006}, and -- as \citet{erwin2005} first
observed -- some that are {\it antitruncated}, with an excess surface brightness relative to an
exponential profile fitted to the inner disk
\citep{erwin2005,pohlen2006,pohlen2007,erwin2007}.  These extended
stellar disks dominate the light past 4--6 scale lengths, and have
flatter profiles with scale lengths that are $\sim50\%$ larger than
the inner disk.

A number of authors have proposed theoretical explanations for
truncated disks.  Dynamical arguments for star--formation thresholds
have been successful in motivating the locations of truncations
\citep{kennicutt1989,schaye2004,naab2006}.  However, observations of
ultraviolet emission
\citep{thilker2005,GildePaz2005,GildePaz2006,boissier2006}, young
stellar populations \citep{Cuillandre2001}, and HII regions
\citep{Ferguson1998} in extended disks suggest that there is indeed
star formation occurring beyond implied thresholds.
More recently, star formation models including a
variety of triggering mechanisms \citep[gravitational instabilities, spiral wave shocks, and stellar and turbulent compression;][]{Elmegreen2006} and N--body
simulations of angular momentum redistribution via bar instabilities 
\citep{Debattista2006} have been more successful at explaining
classically truncated disks. Despite these successes, theoretical mechanisms
 for producing antitruncated disks have received comparatively little attention.

While it is clear that secular processes can influence disk structure
and may produce truncated stellar disks, it has also become apparent
that disk galaxies exist within a hierarchical universe, in
which mergers are a frequent occurrence \citep{laceycole1993,somerville1999,
somerville2000}.  Furthermore, these mergers are likely to play an important
role in shaping the appearance of galaxies.
This is certainly true of collisions
between spiral galaxies of equal mass, so called major mergers, which have
been suggested as the dominant formation mechanism for present-day elliptical
galaxies \citep{toomre1972,toomre1977,negroponte1983,barnes1992,hernquist1992,
hernquist1993a,silk1993,naab2003,robertson2006b,robertson2006a,coxdutta2006}.

There has also been considerable study of the effects of minor mergers
($M_{prim}/M_{sec} \gsim 3$) on the vertical structure and dynamics of
stellar disks \citep{quinn1986,quinn1993,walker1996,huang1997,
sellwood1998,velazquez1999,font2001,ardi2003,brook2004,brook2005,brook2006,
gauthier2006,hayashi2006,kazantzidis2007},
in addition to observational evidence for past interactions with
satellites as the origin of the Milky Way's thick disk
\citep{freeman2002,gilmore2002,wyse2006}.  Moreover, tidal
structures indicative of recent minor mergers have been observed in
both the Milky Way \citep{newberg2002,ibata2003} and M31
\citep{ibata2001,mcconnachie2003}.

In this work, we explore the effects of minor mergers on the structure
of stellar disks at large radius and find that, under certain conditions,
this provides a viable physical mechanism
for producing antitruncated disks.  We demonstrate this process using a
set of hydrodynamical simulations, which are described in \S~\ref{sec:sims}.
An overview of the merger process and specifically the dynamical response of the
stellar disk during a minor merger is provided in \S~\ref{sec:resp}.
\S~\ref{sec:fits} summarizes the surface density profile fitting
procedure, and \S~\ref{sec:gf} and \ref{sec:param} describe the dependence on
parameters of the interaction.  In \S~\ref{sec:discuss} we discuss our results
in comparison to observations of antitruncated disks and within the context of
hierarchical galaxy formation.  Finally, we conclude in \S~\ref{sec:conclude}.

\section{The Simulations}
\label{sec:sims}

For this study, we consider the effects of a 1:8 merger on the
stellar surface density of the primary component's
disk.  These interactions are both cosmologically common
\citep[see, e.g.,][]{laceycole1993,somerville1999,somerville2000} and
kinematically important enough to play a significant role in
determining the appearance of most present--day stellar disks, while at
the same time largely preserving the overall disk structure \citep[see,
e.g.,][]{quinn1993,walker1996,velazquez1999,font2001,kazantzidis2007}.

We consider the idealized case of an isolated interaction, in contrast
to much work on disk galaxy formation done in a full cosmological
context
\citep{font2001,ardi2003,brook2004,brook2005,gauthier2006,kazantzidis2007}.
The cosmological approach has the relative advantage of a more realistic accretion
history.  However, the isolated interactions analyzed here offer the
alternative benefit of examining the individual effects of a single
encounter, and allow us to efficiently sample the parameter space of interactions.  Furthermore, our approach allows the simulation to be performed at much higher resolution.  This helps capture the dynamical effects of minor mergers on the surface density profile at large radius, where resolution is critical.

\begin{table}
\begin{center}
\caption{Model Galaxy Parameters}
\begin{tabular}{ccc}
\hline
\hline
& {\sc Sb} & {\sc Im} \\
\hline
$M$ ($10^{10}$ \msolarunits) & 95 & 12 \\
$V_c$ (\velocityunits) & 160 & 80 \\
$c$ & 9 & 12 \\
$h_{D}$ (\lengthunits) & 4.1 & 1.6 \\
$M_D/M$ & 0.05 & 0.05 \\
$N_H$ & $1.0\times 10^6$ & $1.3\times 10^5$ \\
$N_B$ & $8\times 10^5$  & $4.3\times 10^4$ \\
\hline
\hline
\end{tabular}
\label{tab:models}
\end{center}
\end{table}

The simulations presented in this study were performed with {\sc
Gadget2} \citep{springel2005}, an N--Body/SPH (Smooth Particle
Hydrodynamics) code using the entropy conserving formalism of
\citet{springelhernquist2002}.  We include the effects of radiative
cooling and star formation, tuned to fit the observed Schmidt Law
\citep{schmidt1959,kennicutt1998}.  We also incorporate a
sub--resolution multi--phase feedback model of the interstellar medium
(ISM)
\citep{springelhernquist2003} -- softened ($q_{EOS} = 0.25$)
such that the mass--weighted ISM temperature is $\sim 10^{4.5}$ -- and
sink particles representing supermassive black holes that can accrete
gas and release isotropic thermal energy to the surrounding medium
\citep{springeldimatteo2005a}.  For further details on the progenitor
galaxy models, we refer to \citet{springeldimatteo2005b}, and to other
work done as part of a larger study of the effects of galaxy
interactions on the formation and evolution of galaxies
\citep{dimatteo2005, hopkinshernquist2005c,
hopkinshernquist2005b,hopkinshernquist2005a,
hopkinshernquist2005d,hopkinsall2006,hopkinsred2006,HopkinsSommerville2006,
Hopkins2007a,
robertson2006b,robertson2006a,CoxDimatteo2006,coxdutta2006,CoxJonsson2006}.

\begin{table}
\begin{center}
\caption{Simulation Orbital Parameters}
\begin{tabular}{ccccc}
\hline
\hline
 Name & $i$ & $f_g$ & $R_p$  \\
             &        &            & (\lengthunits) \\ 
\hline
\hline 
{\sc Sb0Im30Rp1} & $30^\circ$ & 0.0 & 5.0 \\
{\sc Sb2Im30Rp1} & $30^\circ$ &  0.2 & 5.0 \\
{\sc Sb4Im30Rp1} & $30^\circ$ & 0.4 & 5.0 \\
{\sc Sb8Im30Rp1} & $30^\circ$ & 0.8 & 5.0 \\
\hline
 {\sc Sb2Im0Rp1} & $0^\circ$  & 0.2 & 5.0 \\
 {\sc Sb2Im90Rp1} & $90^\circ$ &  0.2 & 5.0 \\
 {\sc Sb2Im150Rp1} & $150^\circ$ & 0.2 & 5.0 \\
 {\sc Sb2Im180Rp1} & $180^\circ$ & 0.2 & 5.0 \\
\hline
{\sc Sb2Im30Rp2} & $30^\circ$ & 0.2 & 2.5 \\
{\sc Sb2Im30Rp3} & $30^\circ$ & 0.2 & 10.0 \\
\hline
\hline
\end{tabular}
\label{tab:sims}
\end{center}
\end{table}

A summary of the galaxy models used in the simulations is provided in
Table~\ref{tab:models}, including the total (baryons and dark matter)
mass $M$, circular rotation velocity $V_c$, concentration parameter
$c$, initial disk scale length $h_D$, disk (stars and gas) mass
fraction $M_D/M$, number of dark matter particles in the halo $N_H$,
and baryonic (stars and gas) particles $N_B$.  Both are designed to be
representative of their eponymous local hubble types \citep[see
e.g.,][]{roberts1994}.

The different encounter configurations
considered are summarized in Table~\ref{tab:sims}. 
We assume zero--energy parabolic orbits
($e=1$), as motivated by cosmological simulations
\citep{benson2005,khochfar2006}, with radius of pericenter $R_p$,
orbital inclination ($i$), and primary disk gas fraction $f_g$.  For
the primary disk ({\sc Sb}), we consider 20\% gas disks ($f_g=0.2$),
which are intended to be representative of disks in the local
universe \citep{mcgaugh1997,bell2000}, and higher gas fractions
($f_g=0.4,0.8$) which are consistent with both the more gas--rich local
systems \citep{mcgaugh1997} and high redshift ($z\sim 2$) spirals
\citep{erb2006a}.  The secondary disk has a fixed gas fraction of 40\%
($f_g=0.4$) -- with the exception of the purely collisionless
interaction {\sc Sb0Im30Rp1} which has $f_g=0$ -- and is consistent
with observations of dwarf galaxies and low--mass disks in the local
universe \citep{schombert2001,geha2006}.  The initial spins of the two disks
are not aligned, and are the same in all our simulations.

\section{Dynamical Response of the Stellar Disk to a Minor Merger}
\label{sec:resp}

We find that minor mergers can create antitruncated stellar disks in
face--on spirals.  To examine this effect in detail and
illustrate some of the generic features of a minor merger, we
concentrate on {\sc Sb2Im30Rp1} and its collisionless counterpart {\sc
Sb0Im30Rp1}.  In all cases, the stellar mass surface density profiles 
of the remnant are measured 1 \timeunits after the final coalescence -- or several orbital periods at the half mass radius -- to allow the remnant disk to reach a state of approximate dynamical equilibrium.

In our experiments, antitruncations are produced only when there is a
significant supply of gas in the primary disk.  The driving physical mechanism for
producing this outcome represents a competition between merger
driven inflows and transfer of angular momentum to large radius in the
remnant stellar disk; i.e., gas moves inwards while stars move outwards.  This
effect is manifest in the changes induced in both the gravitational
potential and angular momentum profile during the encounter.

\begin{figure}
\plotone{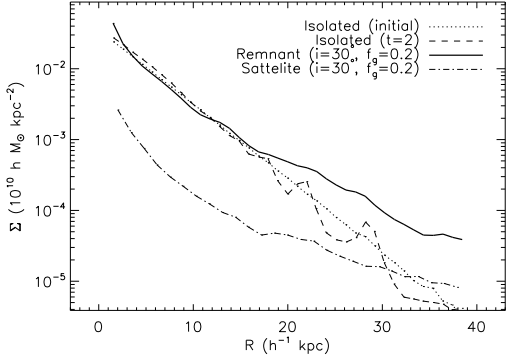}
\caption{Stellar surface density profiles ($\Sigma$) as a function of radial distance from the stellar center of mass ($R$) for an isolated 20\% gas disk {\sc Sb2}, and the remnant {\sc Sb2Im30Rp1}.  Included are the initial {\sc Sb} disk (dotted), an {\sc Sb2} disk evolved in isolation for 2 \timeunits (dashed), and the {\sc Sb2Im30Rp1} remnant (solid).  Also shown is contribution of stars from the secondary disk to the total stellar surface density in the remnant (dash--dot).}
\label{fig:min30_prof}
\end{figure}

In Figure~\ref{fig:min30_prof}, we present the stellar mass surface
density profiles for {\sc Sb2Im30Rp1}, compared to both the initial
primary disk {\sc Sb2} and the same initial disk evolved in isolation
for 2 \timeunits, as a function of radial distance from the stellar
center of mass $R$.  The primary disk is stable; when it is evolved in
isolation over several orbital periods, the inner scale length is only
marginally shorter owing to preferential star formation occurring near
the center \citep[SFR $\sim \rho_g^{1.4}$:][]{kennicutt1998}.  At
large radius, there is some fluctuation of the evolved, isolated disk
about the initial stellar mass surface density.  This owes to Poisson
noise arising from low particle counts at large $R$ and the
development of spiral structure from numerical noise associated with
the discretized dark matter halo \citep{hernquist1993}.

The surface density profile of the {\sc Sb2Im30Rp1} merger remnant shown
in Figure~\ref{fig:min30_prof} displays three key features.  First, within
$4$ \lengthunits, the surface density profile is steep, indicative of
a bulge component produced by the merger.  Second, the surface density
profile from $4 - 20$ \lengthunits is nearly identical to the primary disk.
Third, beyond $20$ \lengthunits, there is a clear excess of surface density
relative to the initial disk of the primary.

\begin{figure}
\plotone{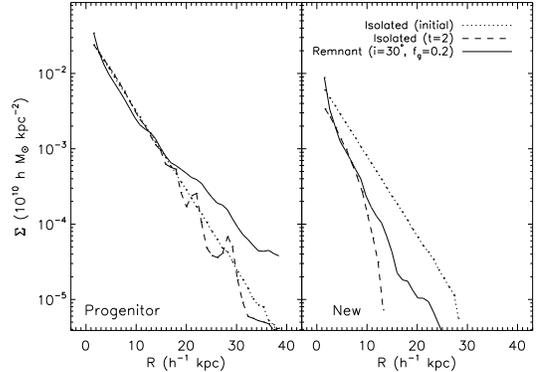}
\caption{Same as Figure~\ref{fig:min30_prof}, but split into the two component stellar populations: ``progenitor'' stars which are present at the start of the simulation, and ``new'' stars which are formed during the interaction.  The dotted line in the right panel shows the initial gas distribution.}
\label{fig:min30_prof_pops}
\end{figure}

The excess surface density in the outer profile of the remnant {\sc Sb2Im30Rp1},
i.e., the antitruncation of its disk, qualitatively -- and, as we will see in
\S~\ref{sec:gf} and \S~\ref{sec:param}, quantitatively -- resembles
the broken exponentials of \citet{pohlen2006}, with an inner scale
length close to that of the initial disk.  In Figure~\ref{fig:min30_prof_pops}, 
we separate out the ``progenitor''
stellar particles that are initialized with the disks and ``new''
stellar particles formed from the gas during the interaction.  We
find that the antitruncation is dominated by progenitor stars.  An
antitruncated disk in {\sc Sb2Im30Rp1} is produced independent of a
fitted profile; a robust result with respect to any fitting procedure.
Separating out the stellar particles that originate in the secondary
stellar disk (see Figure~\ref{fig:min30_prof}), we find that the
increased outer surface density ($R \gsim 20$ \lengthunits)
is dominated by progenitor stars from the primary disk that have been
transferred to larger radius by the interaction.  Furthermore, we find
that these large $R$ features in the profile are rotationally
supported -- their median circular velocity in circular annuli is
$\sim 0.8-0.9$ times the Keplerian orbital velocity at that radius --
and thus long--lived.  We confirm this by evolving our remnant in
isolation for $\sim 10^{10}$ years, and find that the antitruncation
is not a transient feature.

\begin{figure} 
\plotone{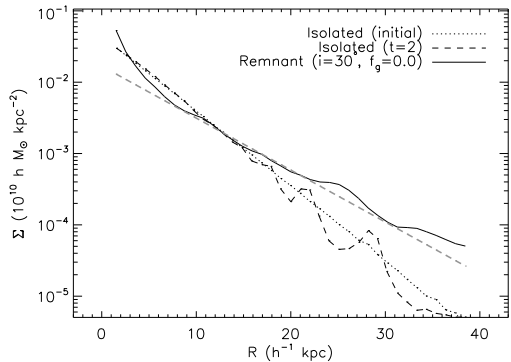}
\caption{Stellar surface density profiles ($\Sigma$) as a function of radial distance from the stellar center of mass ($R$) for an isolated collisionless disk {\sc Sb0}, and the collisionless remnant {\sc Sb0Im30Rp1}.  Included are the initial {\sc Sb} disk (dotted), an {\sc Sb0} disk evolved in isolation for 2 \timeunits (dashed), and the {\sc Sb0Im30Rp1} remnant (solid).  The grey dashed line is an exponential fit to the collisionless remnant {\sc Sb0Im30Rp1} for $R > 10$ \lengthunits, meant to highlight that this profile, in contrast to the 20\% gas remnant  {\sc Sb2Im30Rp1} (see Figure~\ref{fig:min30_prof}), would 
not likely be observed as significantly antitruncated.} 
\label{fig:min30_prof_nocoll}
\end{figure}

The surface density profiles shown in Figure~\ref{fig:min30_prof_nocoll}
demonstrate that the collisionless interaction {\sc Sb0Im30Rp1} displays no
antitruncation in the stellar mass surface density profile of its remnant.
Rather, its surface density profile,
which has been tilted, increasing the scale length at all
radii, does not have a well--defined break. Features at large $R$ are, as in the previous case, also rotationally supported.

Although there is some evidence for bulge formation in {\sc
Sb2Im30Rp1} at $R\lsim 2$ \lengthunits, {\sc Sb0Im30Rp1} shows a much
more pronounced bulge which remains prominent out to larger radii.
This is expected; phase space conservation in a collisionless
interaction leads to lower phase space densities in the core of the
collisionless remnant, and accordingly a more diffuse spheroid
component \citep{hsh1993}.  While bulge growth via minor mergers is a
topic worthy of further study, we postpone a detailed analysis to
future work and point out the qualitative difference between the
surface density profiles of {\sc Sb2Im30Rp1} and {\sc Sb0Im30Rp1}.

\begin{figure}
\plotone{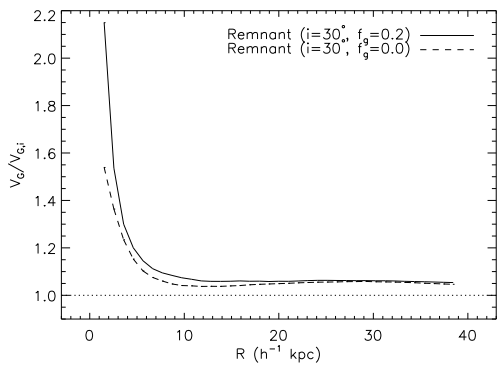}
\caption{Gravitational potential relative to the initial disk, as a function of radial distance from the stellar center of mass ($R$), for the {\sc Sb2Im30Rp1} remnant (solid) and the collisionless {\sc Sb0Im30Rp1} remnant (dashed).  We find that dissipation efficiently deepens the central potential of the remnant relative to the initial disk.}
\label{fig:min30_gravpot}
\end{figure}

To investigate the physical processes driving the antitruncation in
our simulations, we first consider the gravitational potential as a
function of radial distance from the stellar center of mass of
 {\sc Sb0Im30Rp1}, and {\sc Sb2Im30Rp1} as compared to the initial disk in
Figure~\ref{fig:min30_gravpot}.  For certain orbits, a minor merger
can drive nuclear inflows of gas
\citep{hernquist1989,mihos1994,mihos1995}, fueling a centrally
concentrated starburst and creating a deeper potential well there
owing to the effects of gas disspation (see
Figure~\ref{fig:min30_prof_pops}).  This deeper potential will
contract the remnant profile, counteracting the broadening of the profile 
owing to 
angular momentum transfer and maintaining an inner scale length
similar to the initial primary disk.  Furthermore, the newly formed stars will be more
concentrated than the progenitor stars (see
Figure~\ref{fig:min30_prof_pops}), which will also tend to contract
the inner scale length and populate the stellar mass surface density
at small $R$.

\begin{figure}
\plotone{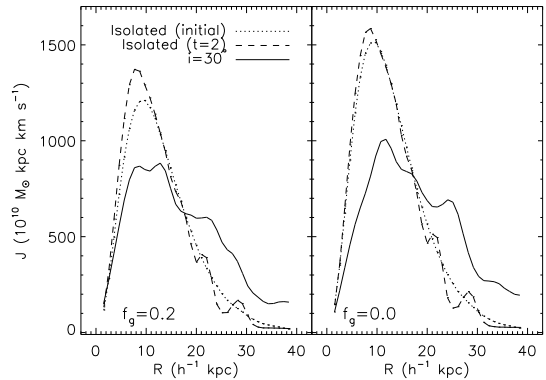}
\caption{Total angular momentum in the stellar disk, as a function of radial distance from the stellar center of mass ($R$).  Shown are: the initial disk (dotted), an isolated disk evolved for 2 \timeunits (dashed), and (solid) the remnant for both a gas fraction of $f_g=0.2$ (left: {\sc Sb2} disk and {\sc Sb2Im30Rp1} remnant) and a collisionless interaction (right: {\sc Sb0} disk and {\sc Sb0Im30Rp1} remnant).  We find that angular momentum is transferred to the outer disk, which increases its scale length relative to the inner disk, and creates the antitruncation.}
\label{fig:min30_ltot}
\end{figure}

At the same time, the interaction transfers angular momentum and
stellar mass to the outer disk.  In Figure~\ref{fig:min30_ltot}, we
show the total angular momentum in circular annuli as a function of
$R$.  In both {\sc Sb2Im30Rp1} and {\sc Sb0Im30Rp1}, the angular
momentum at large $R$ is nearly double that of the initial and evolved
disks; this transfer occurs whether or not gas is included.  In {\sc
Sb2Im30Rp1}, the magnitude of the angular momentum 
in shells at $R < 10$ \lengthunits is also higher, owing to the
more efficient inflows generated by dissipation during the
interaction.  In the collisionless case, because the inner potential
is not as deep, the disk expands more uniformly in response to this
transfer, and therefore does not show an antitruncation.  Thus,
the antitruncation at large $R$ results from expansion of the outer
disk -- similar to that first noted by \citet{quinn1993} -- in response to a net transfer of angular momentum.  When gas is present, the inner potential is deep enough to contract this inner
profile and maintain an inner scale length similar to that of the initial primary disk.

\section{Fitting the Surface Density Profile}
\label{sec:fits}

In \S~\ref{sec:resp}, we find antitruncated stellar disks independent
of the fitted surface density profile.  However, to facilitate
comparison to the observational constraints on antitruncated disks, we
fit profiles to the stellar mass density profiles of our remnants,
projected face--on, and include both progenitor and new stars.  The
fit is performed, as in \S~\ref{sec:resp}, 1 \timeunits after the
final coalescence, so the remnant reaches a state of approximate
dynamical equilibrium.  

Following \citet{pohlen2006}, we mask out the
inner 5 \lengthunits of the disk and first fit an exponential profile
\citep{patterson1940,freeman1970} with scale length $h_{D1}$ to the
inner disk, then fit a second exponential profile with scale length
$h_{D2}$ to the outer disk\footnote{The fitting range for the outer disk was
set by an initial guess for $R_{br}$.  We found that the fitted value for
$R_{br}$ and the outer disk parameters were largely insensitive to this choice.}.  
The break radius $R_{br}$ is defined at
the intersection of the inner and outer disk profiles.  The fits are
performed using a Levenberg-Marquardt least-squares minimization
routine, with bins weighted by the Poisson error ($\sigma_{i} \sim
\Sigma_{i}$).  The results are tabulated in Table~\ref{tab:fits}.  We
note as a caveat that the values of the parameters in our fits are
somewhat sensitive to the manner in which the bins are weighted.
However, using slightly different weights, such as ``flux'' ($\Sigma
r^2$), does not qualitatively affect our results.

Since we are considering the stellar mass distribution out to large
radius, resolution effects are particularly important.  We
note that all of our simulations have more than $1.5\times 10^{4}$
stellar particles in the ``outer'' disk ($4h_{D1} < R < 10h_{D1}$),
and a majority have $\gsim 3.0\times 10^{4}$ over the same
range. Furthermore, tripling the number of particles did not change
the surface density profile at large radius ($R \gsim 20$
\lengthunits) by more than 15\%.  Therefore, we find that our
resolution is sufficient to make robust claims about the the stellar
surface mass density profiles at large radius.

\begin{table}
\begin{center}
\caption{Fitted Surface Density Profiles}
\begin{tabular}{ccccc}
\hline
\hline
 Name & $h_{D1}$ &  $h_{D2}/h_{D1}$ & $R_{br}/h_{D1}$  \\
             &  (\lengthunits) & (\lengthunits) &  &   \\ 
\hline
\hline 
{\sc Sb0Im30Rp1} & 5.99 & \nodata & \nodata \\
{\sc Sb2Im30Rp1} & 4.00 & 1.69 & 3.27 \\
{\sc Sb4Im30Rp1} & 3.54 & 1.85 & 3.33 \\
{\sc Sb8Im30Rp1} & 3.95 & 1.26 & 3.85 \\
\hline
 {\sc Sb2Im0Rp1} & 3.38 & 1.72 & 2.47 \\
 {\sc Sb2Im90Rp1} & 4.10 & 1.24 & 2.64 \\
 {\sc Sb2Im150Rp1} & 4.64 & 1.61 & 5.45 \\
 {\sc Sb2Im180Rp1} & 4.22 & 1.27 & 3.49 \\
\hline
{\sc Sb2Im30Rp2} & 3.70 & 1.80 & 3.10 \\
{\sc Sb2Im30Rp3} & 4.28 & 1.61 & 2.87 \\ 
\hline
\hline
\end{tabular}
\label{tab:fits}
\end{center}
\end{table}

\section{Dependence on the Gas Content}
\label{sec:gf}

\begin{figure*}
\plotone{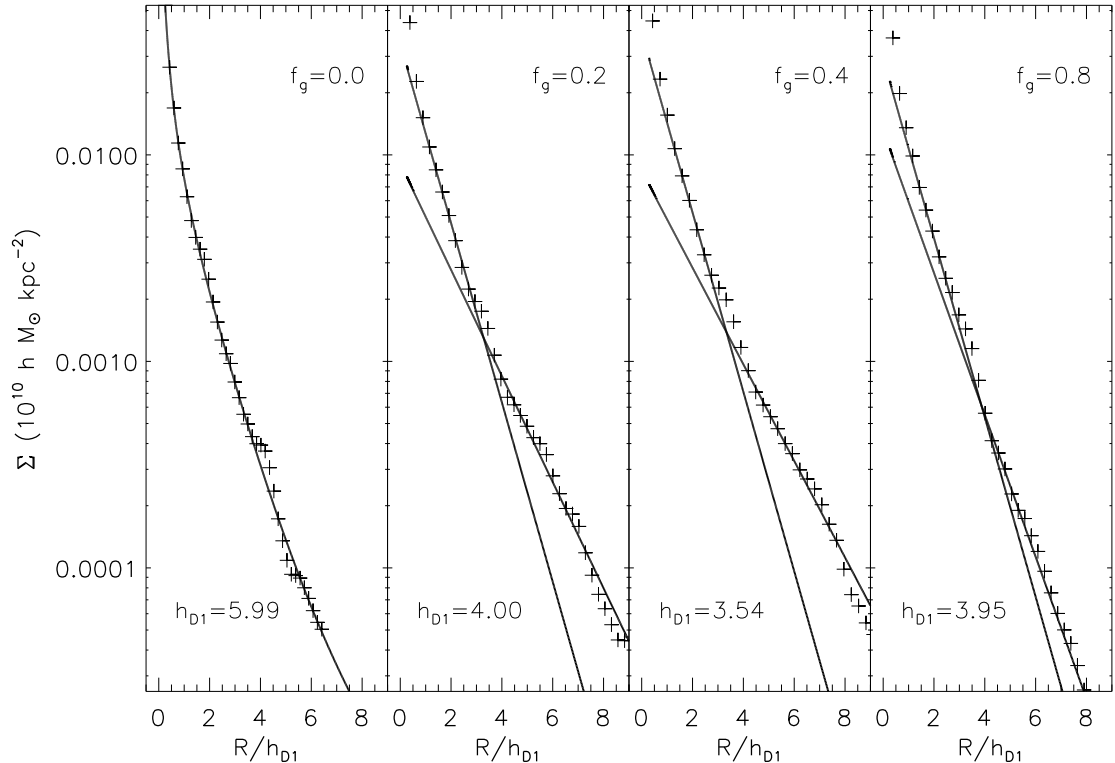}
\caption{Stellar mass surface density of the remnant as a function of radial distance from the center of mass $R$ in units of the inner scale length $h_{D1}$ (see Table~\ref{tab:fits} for fitted values), and its dependence on the initial gas fraction of the primary disk $f_g$: from left to right $f_g = $ 0.0 ({\sc Sb0Im30Rp1}), 0.2 ({\sc Sb2Im30Rp1}), 0.4 ({\sc Sb4Im30Rp1}), and 0.8 ({\sc Sb8Im30Rp1}).  We include both the binned simulation data (crosses), and fitted disk profiles (solid line).  The collisionless ($f_g = 0.0$) remnant is not well--described by a double exponential, and instead has been fit with a combination disk and \citet{deVaucouleurs1959} profile.}
\label{fig:gf_scale}
\end{figure*}

Because antitruncation appears to be a dissipational effect, we expect
the degree and location of the break to depend on the gas content of
the primary disk.  Therefore, we perform a set of experiments varying
the gas fraction of the primary disk, while holding the orbital
parameters fixed.  In Figure~\ref{fig:gf_scale}, we present the
stellar surface mass density profiles of the remnant, including both
the progenitor and new stellar particles, for the four different gas
fractions listed in \S~\ref{sec:sims}.  In addition, we show the
broken exponential disk profiles listed in Table~\ref{tab:fits}.

We find that {\sc Sb0Im30Rp1} -- the collisionless interaction, see
also Figure~\ref{fig:min30_prof_nocoll} -- is not well--fitted by a
broken exponential as in \citet{erwin2005} and \citet{pohlen2006}, and therefore would not
be observed to be antitruncated.  Rather, the scale length increases
 to $h_{D1} \approx 6$\lengthunits, relative to $h_{D} = 4.14$
\lengthunits initially, with a substantial bulge--like component following
an $R^{1/4}$ \citep{deVaucouleurs1959} profile, which is shown in the
fit presented in Figure~\ref{fig:gf_scale}.

\begin{figure}
\plotone{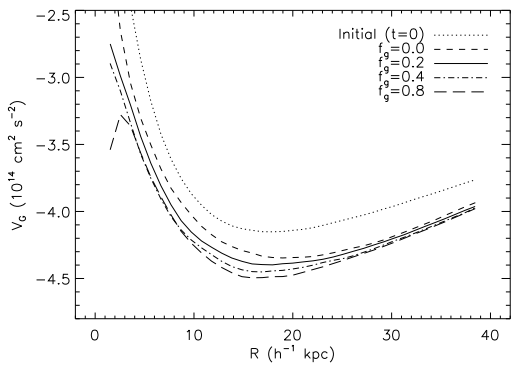}
\caption{Response of the gravitational potential to a minor merger as a function of radial distance from the stellar center of mass $R$, and its dependence on the initial gas fraction of the primary disk $f_g$: $f_g = $ 0.0 (dashed, {\sc Sb0Im30Rp1}), 0.2 (solid, {\sc Sb2Im30Rp1}), 0.4 (dash--dot, {\sc Sb4Im30Rp1}), and 0.8 (long dash, {\sc Sb8Im30Rp1}).  The dotted line shows the initial potential of the primary disk.}
\label{fig:gf_pot}
\end{figure}

Figure~\ref{fig:gf_pot} shows 
that the inner potential is deeper -- and therefore
the inner scale length shorter -- for increased gas fractions (see
Table~\ref{tab:fits}).  The antitruncation is strongest at $f_g=0.2$
and 0.4, while for the highest gas fraction $f_g=0.8$, the relative scale
length $h_{D1}/h_{D2}$ is significantly flatter, resulting in a less
pronounced break.  This is related to the composition of the initial
disk; very gas rich disks have fewer stars remaining in the outer disk
after the initial inflow of gas.  Therefore, as the outer disk
expands, less stellar mass resides at large radii.

\section{Dependence on the Orbital Parameters}
\label{sec:param}

Since the processes that create the antitruncation are dynamical, 
the degree and location of the break should be sensitive to
the orbital parameters of the encounter.  To investigate this,
we vary the orbital parameters -- the orbital inclination
$i$ and radius of pericenter $R_p$ -- holding the gas content of the
primary disk fixed.

\subsection{Orbital Angular Momentum}
\label{sec:lorbit}

To test the ability of the deeper potential to mitigate against the
expansion of the inner disk owing to angular momentum transfer,
we vary the total angular momentum of the secondary's orbit, while
fixing the orbital inclination.  This is done by adjusting the location of 
the pericenter of the secondary's orbit.

\begin{figure*}
\plotone{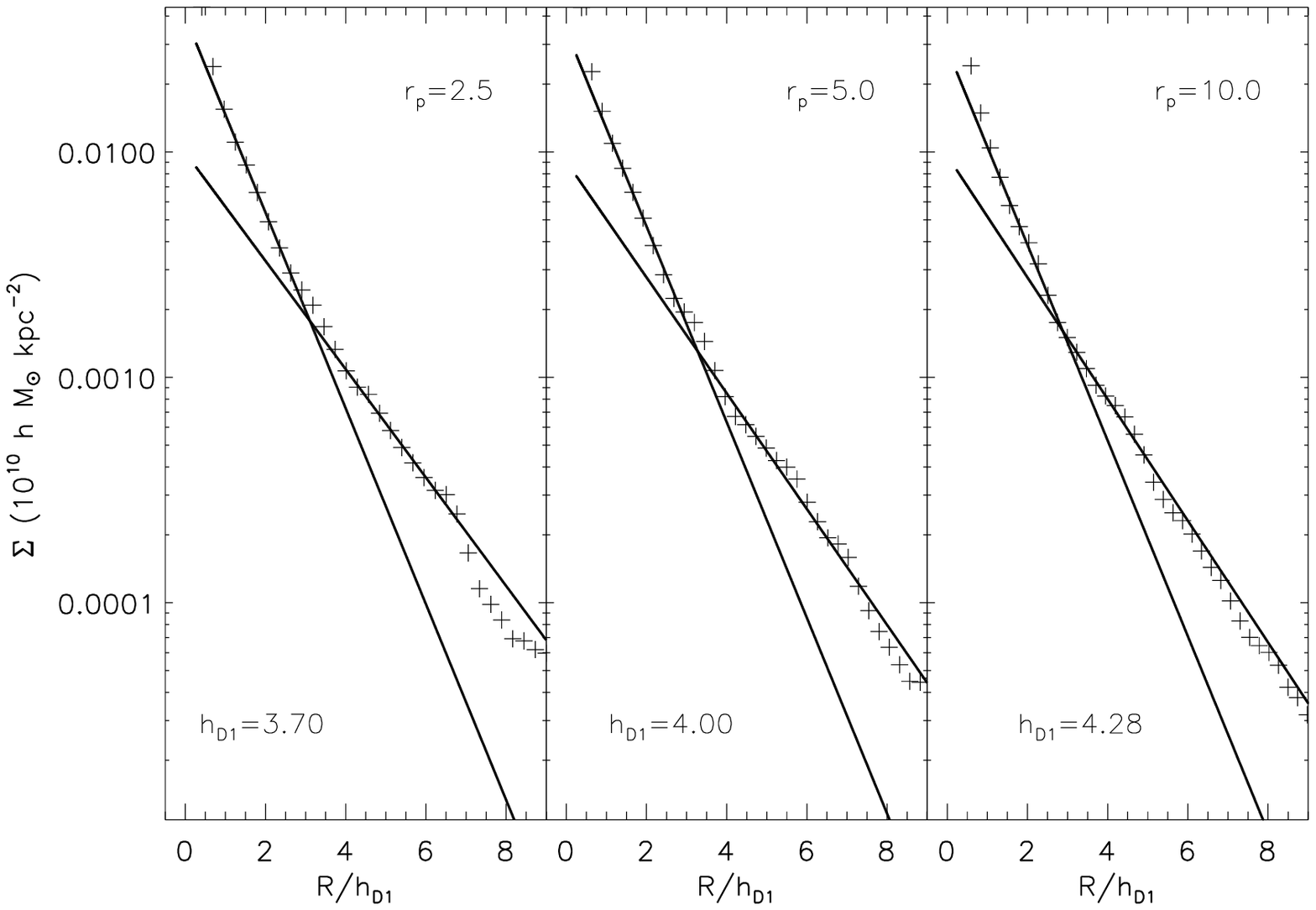}
\caption{Stellar surface mass density of the remnant as a function of radial distance from the center of mass $R$ in units of the inner scale length $h_{D1}$ (see Table~\ref{tab:fits} for fitted values), and its dependence on the radius at pericenter $R_p$ of the orbit of the secondary: from left to right $R_p = $ 2.5 ({\sc Sb2Im30Rp2}), 5.0 ({\sc Sb2Im30Rp1}), and 10.0 ({\sc Sb2Im30Rp3}) \lengthunits.  We include both the binned simulation data (crosses), and fitted disk profiles (solid line).}
\label{fig:rp_scale}
\end{figure*}

\begin{figure}
\plotone{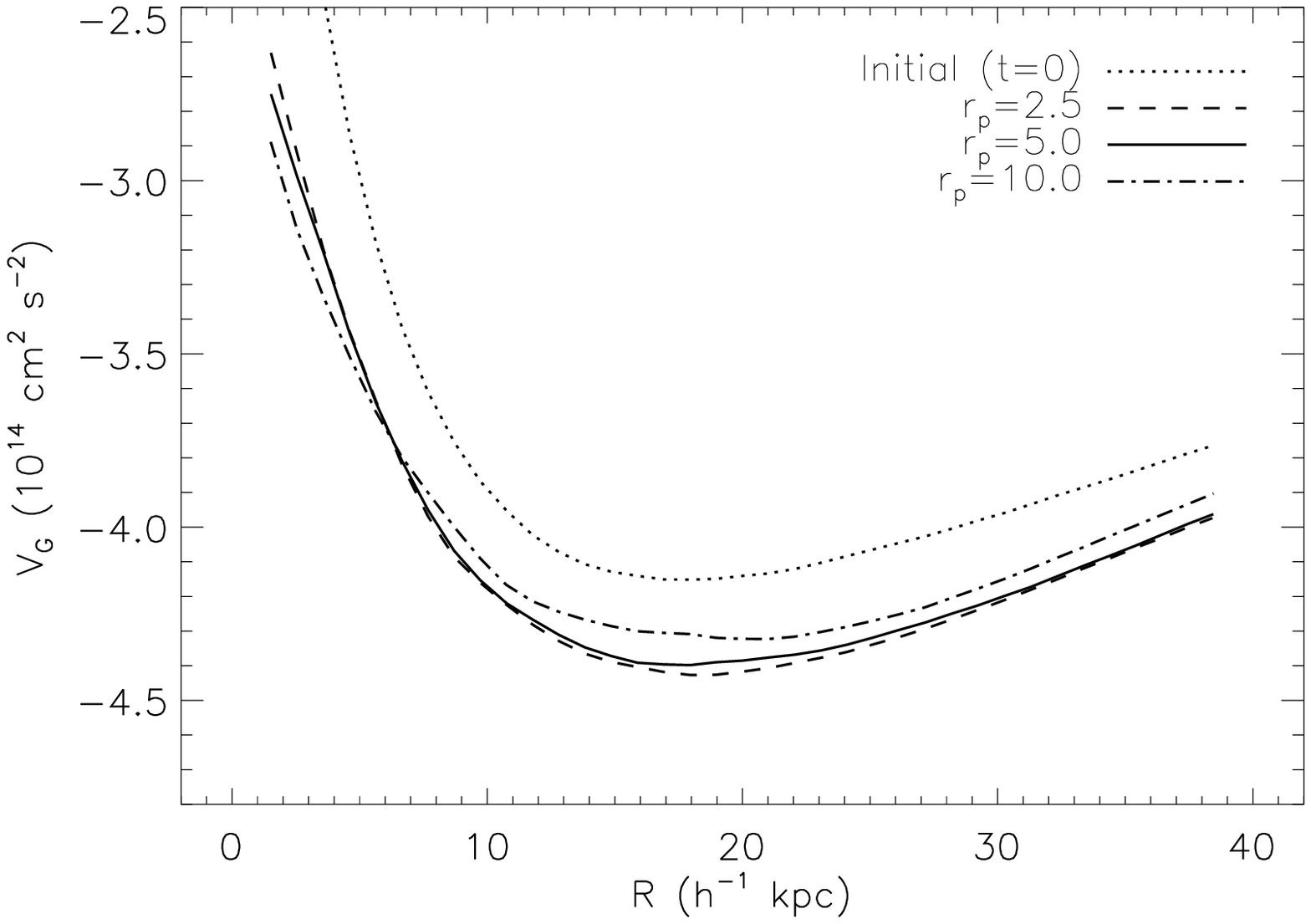}
\caption{Response of the gravitational potential to a minor merger as a function of radial distance from the center of mass $R$, and its dependence on the radius at pericenter $R_p$ of the orbit of the secondary: $R_p = $ 2.5 (dashed, {\sc Sb2Im30Rp2}), 5.0 (solid, {\sc Sb2Im30Rp1}), and 10.0 (dot--dash, {\sc Sb2Im30Rp3}) \lengthunits.}
\label{fig:rp_pot}
\end{figure}

We find that the antitruncation is largely insensitive to increasing the
orbital angular momentum.  Figure~\ref{fig:rp_scale} shows the surface
density profiles for three different radii of pericenter ($R_p$)
spanning a factor of four in orbital angular momentum.  The broken exponential fits
are listed in Table~\ref{tab:fits}. Though the inner scale length does increase
over this range -- despite a deeper inner potential (see
Figure~\ref{fig:rp_pot}) -- it does so by less than 20\%.  Over the same
range, the break radius and relative scale lengths decrease by less
than 10\%.  Therefore, we expect that at fixed inclination, most
orbits would create similar antitruncations.

\subsection{Orbital Inclination}
\label{sec:iorbit}

\begin{figure*}
\plotone{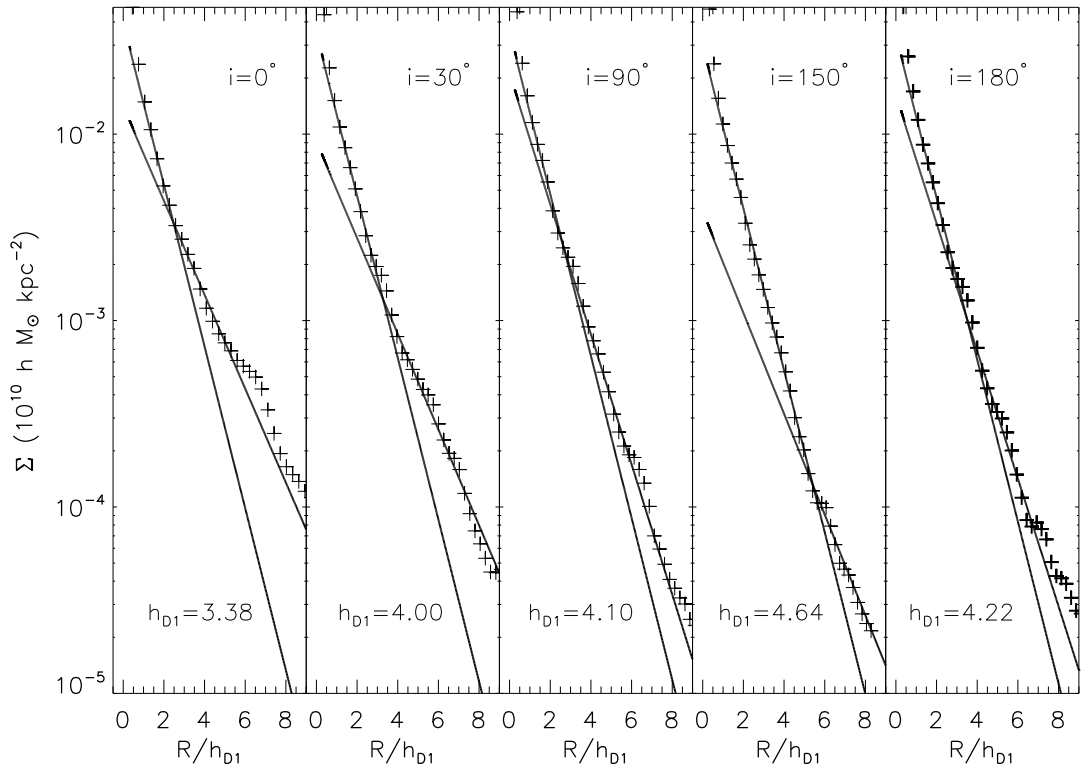}
\caption{Stellar surface mass density of the remnant as a function of radial distance from the center of mass $R$ in units of the inner scale length $h_{D1}$ (see Table~\ref{tab:fits} for fitted values), and its dependence on the orbital inclination $i$ of the interaction: from left to right $i = $ 0 (coplanar prograde, {\sc Sb2Im0Rp1}), 30 (prograde, {\sc Sb2Im30Rp1}), 90 (polar, {\sc Sb2Im90Rp1}), 150 (retrograde, {\sc Sb2Im150Rp1}), and 180 (coplanar retrograde, {\sc Sb2Im180Rp1}).  We include both the binned simulation data (crosses), and fitted disk profiles (solid line).}
\label{fig:inc_scale}
\end{figure*}

\begin{figure}
\plotone{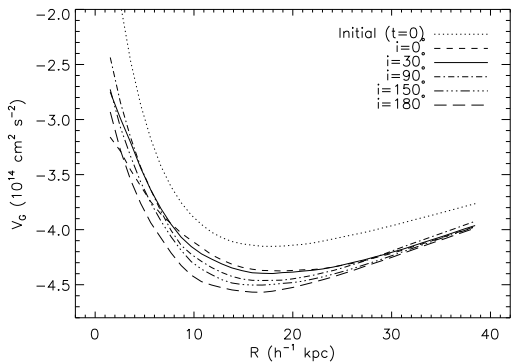}
\caption{Response of the gravitational potential to a minor merger as a function of radial distance from the center of mass $R$, and its dependence on the orbital inclination $i$ of the interaction: $i = $ 0 (coplanar prograde, dashed, {\sc Sb2Im0Rp1}), 30 (prograde, solid, {\sc Sb2Im30Rp1}), 90 (polar, dot--dash, {\sc Sb2Im90Rp1}), 150 (retrograde, double dot--dash, {\sc Sb2Im150Rp1}), and 180 (coplanar retrograde, long dash, {\sc Sb2Im180Rp1}).  The dotted line represents the initial potential of the primary disk.}
\label{fig:inc_pot}
\end{figure}

Varying the orbital inclination of the interaction introduces two
competing effects.  First, prograde minor mergers are more efficient
than retrograde mergers at coupling to the rotation of the primary
disk.  At the same time, coplanar minor mergers are more efficient at
transferring angular momentum to the stellar orbits, inducing bar
formation, and centrally concentrating gas and stars.  As a result,
the closer the interaction is to coplanar, the deeper the remnant's
inner potential.  We present results for five different inclinations,
as outlined in \S~\ref{sec:sims} and Table~\ref{tab:sims}, in
Figures~\ref{fig:inc_scale} and \ref{fig:inc_pot}.

We first consider the prograde and polar interactions -- {\sc
Sb0Im0Rp1}, {\sc Sb0Im30Rp1}, and {\sc Sb0Im90Rp1} -- to illustrate
the combined effects of the merger-driven inflow.  As the inclination
increases from $i=0$ (coplanar prograde) to $i=90$ (polar), the
potential at small scale radius is shallower (see
Figure~\ref{fig:inc_pot}).  Accordingly, the inner scale radii of {\sc
Sb0Im0Rp1}, {\sc Sb0Im30Rp1}, and {\sc Sb0Im90Rp1} are successively
larger with flatter relative outer to inner scale lengths
$h_{D2}/h_{D1}$.  As a result, the antitruncation is less pronounced
for increasing $i$.

When the interaction is retrograde, as with {\sc Sb150Im0Rp1} and {\sc
Sb180Im0Rp1}, the orbital frequency of the secondary is out of resonance
with the orbits of particles in the inner disk.  This makes the encounter
much less efficient at transferring angular momentum to the outer disk.  As a 
result, though there are antitruncations in our retrograde experiments, 
they are far less pronounced than those in prograde encounters.  
Therefore, our simulations suggest that prograde minor mergers will be most
effective at producing antitruncations.

\section{Discussion}
\label{sec:discuss}

We find that minor mergers can create antitruncated stellar disks in
face--on spiral galaxies, and that this antitruncation is produced by
a competition between merger driven inflows of gas and transfer of
angular momentum to large $R$ in the remnant stellar disk that moves
stars outwards.  Because this process requires gaseous inflows, 
antitruncated stellar disks are produced only when gas dissipation and star
formation are included.  Moreover, the magnitude and location of
antitruncation is related to both the gas content of the primary disk
and the orbital parameters of the interaction.  These features in the
surface stellar mass density profile are rotationally supported, and
therefore long--lived and likely to be observed in local spirals.

This merger-driven scenario for the production of antitruncated disks
is supported by observations of face--on spirals that find antitruncated
disks occur more frequently in earlier--type spirals and in higher density
environments \citep{pohlen2006}.  The authors note
that the frequency of antitruncated disks
increases from 20\% in Sd types to 50\% in Sb types, while the
fraction of classically truncated disks decreases from 40\% to 10\%
over the same range.  This agrees qualitatively with minor mergers as the 
physical mechanism driving disk antitruncation: spirals in higher density environments are
more likely to have undergone minor mergers which create
systematically earlier Hubble types \citep{naab2003}.  At the same time, both 
\citet{erwin2005} and \citet{pohlen2006} show observational evidence for asymmetries
or recent interactions in antitruncated systems, which further supports a merger--driven scenario.

Using the extended Press-Schechter formalism
\citep{jenkins2001} and the method of \citet{laceycole1993} to
estimate halo merger histories, and assuming the cosmology of
\citet{spergel2003}, we find that {\sc Sb}--type halos ($M_{tot}\sim
10^{12}$ \msolarunits) are likely to experience of order one 1:8 minor
merger from $z=1$ to the present--day
\citep{hopkins2007c,hopkins2007b}.  In our simulations, strong
antitruncations are produced when the orbit of the secondary is
inclined ($0^\circ \lsim i \lsim 90^\circ$), prograde, and has
moderate angular momentum ($R_p \gsim h_{D}$).  So, if all orbits are
distributed isotropically -- i.e., equally likely in bins of
$d\cos{i}$ -- and follow the distribution of $R_p$ as inferred from
cosmological N--body simulations \citep[e.g.,][]{benson2005}, then we
would expect $\sim 40\%-50\%$ of Sb type spirals to have pronounced
antitruncations.

At the same time, our fits agree quantitatively with the relative
scale lengths $h_{D2}/h_{D1}$ observed by \citet{pohlen2006} and \citet{erwin2005}.  We
find, however, that the break radius in our simulations is at the low end of the observed range; 
\citet{pohlen2006} and \citet{erwin2005} find $R_{br}/h_{D1} = 3-6$ 
while in our simulations $R_{br}/h_{D1} = 3-4$.
This may owe either to: (1) the smaller mass of a typical galaxy in the \citet{pohlen2006} sample, or
(2) the limited range of parameter space spanned by our simulations.
The observations of \citet{pohlen2006} appear to be dominated by
somewhat lower--mass spirals than the Milky--Way mass primary disk in the
interactions examined here.  This could potentially lead to shorter inner scale
lengths \citep[e.g.,][]{courteau1996,dejong1996}, which may tend to
increase the average observed $R_{br}/h_{D1}$ ratio.  Also, our
simulations sample only a small subset of the parameter space for
individual interactions.  The real merging history of galaxies
likely involves a variety of mass fractions and multiple mergers
that may produce subtly different effects.  However, we find that our
simulations of minor mergers generically lead to antitruncated disks
for a range of orbital geometries, and therefore represent a viable
mechanism for producing these features.

More locally, \citet{ibata2005} and \citet{ibata2007} recently observed an extended stellar disk in M31.  Though it is not entirely clear whether or not this feature is preceded by a well defined break in the surface brightness profile, it is possible that the extended disk represents an antitruncation of the type observed by \citet{erwin2005} and \citet{pohlen2006}.  \citet{ibata2005} estimate that it contains roughly 10\% of the stellar mass and 30\% of the angular momentum of the total stellar disk, as compared to 5\% of the mass and 45\% of the angular momentum in the antitruncated disk of {\sc Sb2Im30Rp1}.  Furthermore, the kinematics -- specifically, the dispersion of circular velocity lags relative to Keplerian rotation -- of resolved stars in the extended disk show evidence of dynamical heating which could have been caused by a minor merger \citep[e.g.][]{quinn1993,walker1996,velazquez1999}.  Therefore, although we cannot say with certainty that the extended disk of M31 represents a local example of an antitruncation, it is broadly consistent with our modeling.

\section{Conclusion}
\label{sec:conclude}

We use hydrodynamic simulations to investigate minor mergers as a
physical mechanism for creating antitruncated disks in face--on
spirals.  We find that the antitruncation is produced by two competing
effects: merger--driven gas inflows deepen the central potential and
contract the inner profile, while at the same time angular momentum is
transferred to large radius and causes the outer disk to expand.
Because the inflows are far more efficient when gas dissipation is
included, the antitruncation is produced in our experiments only when
a significant gas supply is present in the initial primary disk.  This
effect is also only seen when the interaction is prograde, rather
than polar or retrograde, with moderate ($R_p \sim h_{D}$) orbital
angular momentum.  

Our merger-driven scenario for producing antitruncated disks yields
results that agree with observations of local face--on spirals \citep{erwin2005,pohlen2006}, 
both in terms of the parameters of the antitruncation and its frequency with Hubble type.  Therefore, we find that minor mergers are a viable physical mechanism for producing antitruncated disks.
\newpage
\acknowledgements

Particular thanks to the referee, Michael Pohlen, for his comments and suggestions that improved this manuscript.  Also thanks to Peter Erwin, Yuexing Li, Phillip F. Hopkins, Beth Willman, and Du\v{s}an Kere\v{s} for helpful discussions.  These simulations were performed at the Harvard Institute for Theory and Computation at the Smithsonian Center for Astrophysics.

\bibliographystyle{apj}
\bibliography{../../minors}

\begin{thebibliography}{106}
\expandafter\ifx\csname natexlab\endcsname\relax\def\natexlab#1{#1}\fi

\bibitem[{{Ardi} {et~al.}(2003){Ardi}, {Tsuchiya}, \& {Burkert}}]{ardi2003}
{Ardi}, E., {Tsuchiya}, T., \& {Burkert}, A. 2003, \apj, 596, 204

\bibitem[{{Barnaby} \& {Thronson}(1992)}]{barnaby1992}
{Barnaby}, D. \& {Thronson}, Jr., H.~A. 1992, \aj, 103, 41

\bibitem[{{Barnes}(1992)}]{barnes1992}
{Barnes}, J.~E. 1992, \apj, 393, 484

\bibitem[{{Barteldrees} \& {Dettmar}(1994)}]{barteldrees1994}
{Barteldrees}, A. \& {Dettmar}, R.-J. 1994, \aaps, 103, 475

\bibitem[{{Bell} \& {de Jong}(2000)}]{bell2000}
{Bell}, E.~F. \& {de Jong}, R.~S. 2000, \mnras, 312, 497

\bibitem[{{Benson}(2005)}]{benson2005}
{Benson}, A.~J. 2005, \mnras, 358, 551

\bibitem[{{Bland-Hawthorn} {et~al.}(2005){Bland-Hawthorn}, {Vlaji{\'c}},
  {Freeman}, \& {Draine}}]{BlandHawthorn2005}
{Bland-Hawthorn}, J., {Vlaji{\'c}}, M., {Freeman}, K.~C., \& {Draine}, B.~T.
  2005, \apj, 629, 239

\bibitem[{{Boissier} {et~al.}(2006)}]{boissier2006}
{Boissier}, S. {et~al.} 2006, astro-ph/0609071

\bibitem[{{Brook} {et~al.}(2006){Brook}, {Richard}, {Kawata}, {Martel}, \&
  {Gibson}}]{brook2006}
{Brook}, C., {Richard}, S., {Kawata}, D., {Martel}, H., \& {Gibson}, B.~K.
  2006, astro-ph/0611748

\bibitem[{{Brook} {et~al.}(2005){Brook}, {Gibson}, {Martel}, \&
  {Kawata}}]{brook2005}
{Brook}, C.~B., {Gibson}, B.~K., {Martel}, H., \& {Kawata}, D. 2005, \apj, 630,
  298

\bibitem[{{Brook} {et~al.}(2004){Brook}, {Kawata}, {Gibson}, \&
  {Freeman}}]{brook2004}
{Brook}, C.~B., {Kawata}, D., {Gibson}, B.~K., \& {Freeman}, K.~C. 2004, \apj,
  612, 894

\bibitem[{{Courteau}(1996)}]{courteau1996}
{Courteau}, S. 1996, \apjs, 103, 363

\bibitem[{{Cox} {et~al.}(2006{\natexlab{a}}){Cox}, {Di Matteo}, {Hernquist},
  {Hopkins}, {Robertson}, \& {Springel}}]{CoxDimatteo2006}
{Cox}, T.~J., {Di Matteo}, T., {Hernquist}, L., {Hopkins}, P.~F., {Robertson},
  B., \& {Springel}, V. 2006{\natexlab{a}}, \apj, 643, 692

\bibitem[{{Cox} {et~al.}(2006{\natexlab{b}}){Cox}, {Dutta}, {Di Matteo},
  {Hernquist}, {Hopkins}, {Robertson}, \& {Springel}}]{coxdutta2006}
{Cox}, T.~J., {Dutta}, S.~N., {Di Matteo}, T., {Hernquist}, L., {Hopkins},
  P.~F., {Robertson}, B., \& {Springel}, V. 2006{\natexlab{b}}, \apj, 650, 791

\bibitem[{{Cox} {et~al.}(2006{\natexlab{c}}){Cox}, {Jonsson}, {Primack}, \&
  {Somerville}}]{CoxJonsson2006}
{Cox}, T.~J., {Jonsson}, P., {Primack}, J.~R., \& {Somerville}, R.~S.
  2006{\natexlab{c}}, \mnras, 373, 1013

\bibitem[{{Cuillandre} {et~al.}(2001){Cuillandre}, {Lequeux}, {Allen},
  {Mellier}, \& {Bertin}}]{Cuillandre2001}
{Cuillandre}, J.-C., {Lequeux}, J., {Allen}, R.~J., {Mellier}, Y., \& {Bertin},
  E. 2001, \apj, 554, 190

\bibitem[{{de Grijs} {et~al.}(2001){de Grijs}, {Kregel}, \&
  {Wesson}}]{deGrijs2001}
{de Grijs}, R., {Kregel}, M., \& {Wesson}, K.~H. 2001, \mnras, 324, 1074

\bibitem[{{de Jong}(1996)}]{dejong1996}
{de Jong}, R.~S. 1996, \aap, 313, 45

\bibitem[{{de Vaucouleurs}(1959)}]{deVaucouleurs1959}
{de Vaucouleurs}, G. 1959, Handbuch der Physik, 53, 311

\bibitem[{{Debattista} {et~al.}(2006){Debattista}, {Mayer}, {Carollo}, {Moore},
  {Wadsley}, \& {Quinn}}]{Debattista2006}
{Debattista}, V.~P., {Mayer}, L., {Carollo}, C.~M., {Moore}, B., {Wadsley}, J.,
  \& {Quinn}, T. 2006, \apj, 645, 209

\bibitem[{{Di Matteo} {et~al.}(2005){Di Matteo}, {Springel}, \&
  {Hernquist}}]{dimatteo2005}
{Di Matteo}, T., {Springel}, V., \& {Hernquist}, L. 2005, \nat, 433, 604

\bibitem[{{Elmegreen} \& {Hunter}(2006)}]{Elmegreen2006}
{Elmegreen}, B.~G. \& {Hunter}, D.~A. 2006, \apj, 636, 712

\bibitem[{{Erb} {et~al.}(2006){Erb}, {Steidel}, {Shapley}, {Pettini}, {Reddy},
  \& {Adelberger}}]{erb2006a}
{Erb}, D.~K., {Steidel}, C.~C., {Shapley}, A.~E., {Pettini}, M., {Reddy},
  N.~A., \& {Adelberger}, K.~L. 2006, \apj, 646, 107

\bibitem[{{Erwin} {et~al.}(2005){Erwin}, {Beckman}, \& {Pohlen}}]{erwin2005}
{Erwin}, P., {Beckman}, J.~E., \& {Pohlen}, M. 2005, \apjl, 626, L81

\bibitem[{{Erwin} {et~al.}(2007){Erwin}, {Pohlen}, \& {Beckman}}]{erwin2007}
{Erwin}, P., {Pohlen}, M., \& {Beckman}, J.~E. 2007, AJ, submitted

\bibitem[{{Ferguson} {et~al.}(1998){Ferguson}, {Wyse}, {Gallagher}, \&
  {Hunter}}]{Ferguson1998}
{Ferguson}, A.~M.~N., {Wyse}, R.~F.~G., {Gallagher}, J.~S., \& {Hunter}, D.~A.
  1998, \apjl, 506, L19

\bibitem[{{Florido} {et~al.}(2006{\natexlab{a}}){Florido}, {Battaner},
  {Guijarro}, {Garz{\'o}n}, \& {Castillo-Morales}}]{Florido2006b}
{Florido}, E., {Battaner}, E., {Guijarro}, A., {Garz{\'o}n}, F., \&
  {Castillo-Morales}, A. 2006{\natexlab{a}}, \aap, 455, 475

\bibitem[{{Florido} {et~al.}(2006{\natexlab{b}}){Florido}, {Battaner},
  {Guijarro}, {Garz{\'o}n}, \& {Castillo-Morales}}]{Florido2006a}
---. 2006{\natexlab{b}}, \aap, 455, 467

\bibitem[{{Font} {et~al.}(2001){Font}, {Navarro}, {Stadel}, \&
  {Quinn}}]{font2001}
{Font}, A.~S., {Navarro}, J.~F., {Stadel}, J., \& {Quinn}, T. 2001, \apjl, 563,
  L1

\bibitem[{{Freeman} \& {Bland-Hawthorn}(2002)}]{freeman2002}
{Freeman}, K. \& {Bland-Hawthorn}, J. 2002, \araa, 40, 487

\bibitem[{{Freeman}(1970)}]{freeman1970}
{Freeman}, K.~C. 1970, \apj, 160, 811

\bibitem[{{Gauthier} {et~al.}(2006){Gauthier}, {Dubinski}, \&
  {Widrow}}]{gauthier2006}
{Gauthier}, J.-R., {Dubinski}, J., \& {Widrow}, L.~M. 2006, \apj, 653, 1180

\bibitem[{{Geha} {et~al.}(2006){Geha}, {Blanton}, {Masjedi}, \&
  {West}}]{geha2006}
{Geha}, M., {Blanton}, M.~R., {Masjedi}, M., \& {West}, A.~A. 2006, \apj, 653,
  240

\bibitem[{{Gil de Paz} {et~al.}(2005){Gil de Paz}, {Madore}, {Boissier},
  {Swaters}, {Popescu}, {Tuffs}, {Sheth}, {Kennicutt}, {Bianchi}, {Thilker}, \&
  {Martin}}]{GildePaz2005}
{Gil de Paz}, A., {Madore}, B.~F., {Boissier}, S., {Swaters}, R., {Popescu},
  C.~C., {Tuffs}, R.~J., {Sheth}, K., {Kennicutt}, Jr., R.~C., {Bianchi}, L.,
  {Thilker}, D., \& {Martin}, D.~C. 2005, \apjl, 627, L29

\bibitem[{{Gil de Paz} {et~al.}(2006)}]{GildePaz2006}
{Gil de Paz}, A. {et~al.} 2006, astro-ph/0606440

\bibitem[{{Gilmore} {et~al.}(2002){Gilmore}, {Wyse}, \& {Norris}}]{gilmore2002}
{Gilmore}, G., {Wyse}, R.~F.~G., \& {Norris}, J.~E. 2002, \apjl, 574, L39

\bibitem[{{Hayashi} \& {Chiba}(2006)}]{hayashi2006}
{Hayashi}, H. \& {Chiba}, M. 2006, \pasj, 58, 835

\bibitem[{{Hernquist}(1989)}]{hernquist1989}
{Hernquist}, L. 1989, \nat, 340, 687

\bibitem[{{Hernquist}(1993{\natexlab{a}})}]{hernquist1993}
---. 1993{\natexlab{a}}, \apjs, 86, 389

\bibitem[{{Hernquist}(1993{\natexlab{b}})}]{hernquist1993a}
---. 1993{\natexlab{b}}, \apj, 409, 548

\bibitem[{{Hernquist} \& {Mihos}(1995)}]{mihos1995}
{Hernquist}, L. \& {Mihos}, J.~C. 1995, \apj, 448, 41

\bibitem[{{Hernquist} {et~al.}(1993){Hernquist}, {Spergel}, \&
  {Heyl}}]{hsh1993}
{Hernquist}, L., {Spergel}, D.~N., \& {Heyl}, J.~S. 1993, \apj, 416, 415

\bibitem[{{Hernquist} \& {Weinberg}(1992)}]{hernquist1992}
{Hernquist}, L. \& {Weinberg}, M.~D. 1992, \apj, 400, 80

\bibitem[{{Hopkins} {et~al.}(2007{\natexlab{a}}){Hopkins}, {Cox}, {Keres}, \&
  {Hernquist}}]{hopkins2007c}
{Hopkins}, P.~F., {Cox}, T.~J., {Keres}, D., \& {Hernquist}, L.
  2007{\natexlab{a}}, ApJ, submitted (astro-ph/0706.1246)

\bibitem[{{Hopkins} {et~al.}(2005{\natexlab{a}}){Hopkins}, {Hernquist}, {Cox},
  {Di Matteo}, {Martini}, {Robertson}, \& {Springel}}]{hopkinshernquist2005c}
{Hopkins}, P.~F., {Hernquist}, L., {Cox}, T.~J., {Di Matteo}, T., {Martini},
  P., {Robertson}, B., \& {Springel}, V. 2005{\natexlab{a}}, \apj, 630, 705

\bibitem[{{Hopkins} {et~al.}(2005{\natexlab{b}}){Hopkins}, {Hernquist}, {Cox},
  {Di Matteo}, {Robertson}, \& {Springel}}]{hopkinshernquist2005b}
{Hopkins}, P.~F., {Hernquist}, L., {Cox}, T.~J., {Di Matteo}, T., {Robertson},
  B., \& {Springel}, V. 2005{\natexlab{b}}, \apj, 630, 716

\bibitem[{{Hopkins} {et~al.}(2005{\natexlab{c}}){Hopkins}, {Hernquist}, {Cox},
  {Di Matteo}, {Robertson}, \& {Springel}}]{hopkinshernquist2005a}
---. 2005{\natexlab{c}}, \apj, 632, 81

\bibitem[{{Hopkins} {et~al.}(2006{\natexlab{a}}){Hopkins}, {Hernquist}, {Cox},
  {Di Matteo}, {Robertson}, \& {Springel}}]{hopkinsall2006}
---. 2006{\natexlab{a}}, \apjs, 163, 1

\bibitem[{{Hopkins} {et~al.}(2007{\natexlab{b}}){Hopkins}, {Hernquist}, {Cox},
  \& {Keres}}]{hopkins2007b}
{Hopkins}, P.~F., {Hernquist}, L., {Cox}, T.~J., \& {Keres}, D.
  2007{\natexlab{b}}, ApJ, submitted (astro-ph/0706.1243)

\bibitem[{{Hopkins} {et~al.}(2007{\natexlab{c}}){Hopkins}, {Hernquist}, {Cox},
  {Robertson}, \& {Krause}}]{Hopkins2007a}
{Hopkins}, P.~F., {Hernquist}, L., {Cox}, T.~J., {Robertson}, B., \& {Krause},
  E. 2007{\natexlab{c}}, astro-ph/0701351

\bibitem[{{Hopkins} {et~al.}(2006{\natexlab{b}}){Hopkins}, {Hernquist}, {Cox},
  {Robertson}, \& {Springel}}]{hopkinsred2006}
{Hopkins}, P.~F., {Hernquist}, L., {Cox}, T.~J., {Robertson}, B., \&
  {Springel}, V. 2006{\natexlab{b}}, \apjs, 163, 50

\bibitem[{{Hopkins} {et~al.}(2005{\natexlab{d}}){Hopkins}, {Hernquist},
  {Martini}, {Cox}, {Robertson}, {Di Matteo}, \&
  {Springel}}]{hopkinshernquist2005d}
{Hopkins}, P.~F., {Hernquist}, L., {Martini}, P., {Cox}, T.~J., {Robertson},
  B., {Di Matteo}, T., \& {Springel}, V. 2005{\natexlab{d}}, \apjl, 625, L71

\bibitem[{{Hopkins} {et~al.}(2006{\natexlab{c}}){Hopkins}, {Somerville},
  {Hernquist}, {Cox}, {Robertson}, \& {Li}}]{HopkinsSommerville2006}
{Hopkins}, P.~F., {Somerville}, R.~S., {Hernquist}, L., {Cox}, T.~J.,
  {Robertson}, B., \& {Li}, Y. 2006{\natexlab{c}}, \apj, 652, 864

\bibitem[{{Huang} \& {Carlberg}(1997)}]{huang1997}
{Huang}, S. \& {Carlberg}, R.~G. 1997, \apj, 480, 503

\bibitem[{{Hunter} \& {Elmegreen}(2006)}]{hunter2006}
{Hunter}, D.~A. \& {Elmegreen}, B.~G. 2006, \apjs, 162, 49

\bibitem[{{Ibata} {et~al.}(2005){Ibata}, {Chapman}, {Ferguson}, {Lewis},
  {Irwin}, \& {Tanvir}}]{ibata2005}
{Ibata}, R., {Chapman}, S., {Ferguson}, A.~M.~N., {Lewis}, G., {Irwin}, M., \&
  {Tanvir}, N. 2005, \apj, 634, 287

\bibitem[{{Ibata} {et~al.}(2001){Ibata}, {Irwin}, {Lewis}, {Ferguson}, \&
  {Tanvir}}]{ibata2001}
{Ibata}, R., {Irwin}, M., {Lewis}, G., {Ferguson}, A.~M.~N., \& {Tanvir}, N.
  2001, \nat, 412, 49

\bibitem[{{Ibata} {et~al.}(2007){Ibata}, {Martin}, {Irwin}, {Chapman},
  {Ferguson}, {Lewis}, \& {McConnachie}}]{ibata2007}
{Ibata}, R., {Martin}, N.~F., {Irwin}, M., {Chapman}, S., {Ferguson}, A.~M.~N.,
  {Lewis}, G.~F., \& {McConnachie}, A.~W. 2007, astro-ph/0704.1318

\bibitem[{{Ibata} {et~al.}(2003){Ibata}, {Irwin}, {Lewis}, {Ferguson}, \&
  {Tanvir}}]{ibata2003}
{Ibata}, R.~A., {Irwin}, M.~J., {Lewis}, G.~F., {Ferguson}, A.~M.~N., \&
  {Tanvir}, N. 2003, \mnras, 340, L21

\bibitem[{{Jenkins} {et~al.}(2001){Jenkins}, {Frenk}, {White}, {Colberg},
  {Cole}, {Evrard}, {Couchman}, \& {Yoshida}}]{jenkins2001}
{Jenkins}, A., {Frenk}, C.~S., {White}, S.~D.~M., {Colberg}, J.~M., {Cole}, S.,
  {Evrard}, A.~E., {Couchman}, H.~M.~P., \& {Yoshida}, N. 2001, \mnras, 321,
  372

\bibitem[{{Kazantzidis} {et~al.}(2007)}]{kazantzidis2007}
{Kazantzidis}, S. {et~al.} 2007, in preparation

\bibitem[{{Kennicutt}(1989)}]{kennicutt1989}
{Kennicutt}, Jr., R.~C. 1989, \apj, 344, 685

\bibitem[{{Kennicutt}(1998)}]{kennicutt1998}
---. 1998, \apj, 498, 541

\bibitem[{{Khochfar} \& {Burkert}(2006)}]{khochfar2006}
{Khochfar}, S. \& {Burkert}, A. 2006, \aap, 445, 403

\bibitem[{{Kregel} {et~al.}(2002){Kregel}, {van der Kruit}, \& {de
  Grijs}}]{kregel2002}
{Kregel}, M., {van der Kruit}, P.~C., \& {de Grijs}, R. 2002, \mnras, 334, 646

\bibitem[{{Lacey} \& {Cole}(1993)}]{laceycole1993}
{Lacey}, C. \& {Cole}, S. 1993, \mnras, 262, 627

\bibitem[{{McConnachie} {et~al.}(2003){McConnachie}, {Irwin}, {Ibata},
  {Ferguson}, {Lewis}, \& {Tanvir}}]{mcconnachie2003}
{McConnachie}, A.~W., {Irwin}, M.~J., {Ibata}, R.~A., {Ferguson}, A.~M.~N.,
  {Lewis}, G.~F., \& {Tanvir}, N. 2003, \mnras, 343, 1335

\bibitem[{{McGaugh} \& {de Blok}(1997)}]{mcgaugh1997}
{McGaugh}, S.~S. \& {de Blok}, W.~J.~G. 1997, \apj, 481, 689

\bibitem[{{Mihos} \& {Hernquist}(1994)}]{mihos1994}
{Mihos}, J.~C. \& {Hernquist}, L. 1994, \apjl, 431, L9

\bibitem[{{Naab} \& {Burkert}(2003)}]{naab2003}
{Naab}, T. \& {Burkert}, A. 2003, \apj, 597, 893

\bibitem[{{Naab} \& {Ostriker}(2006)}]{naab2006}
{Naab}, T. \& {Ostriker}, J.~P. 2006, \mnras, 366, 899

\bibitem[{{Negroponte} \& {White}(1983)}]{negroponte1983}
{Negroponte}, J. \& {White}, S.~D.~M. 1983, \mnras, 205, 1009

\bibitem[{{Newberg} {et~al.}(2002)}]{newberg2002}
{Newberg}, H.~J. {et~al.} 2002, \apj, 569, 245

\bibitem[{{Patterson}(1940)}]{patterson1940}
{Patterson}, F.~S. 1940, Harvard College Observatory Bulletin, 914, 9

\bibitem[{{P{\'e}rez}(2004)}]{perez2004}
{P{\'e}rez}, I. 2004, \aap, 427, L17

\bibitem[{{Pohlen} {et~al.}(2000){Pohlen}, {Dettmar}, \&
  {L{\"u}tticke}}]{pohlen2000}
{Pohlen}, M., {Dettmar}, R.-J., \& {L{\"u}tticke}, R. 2000, \aap, 357, L1

\bibitem[{{Pohlen} {et~al.}(2002){Pohlen}, {Dettmar}, {L{\"u}tticke}, \&
  {Aronica}}]{pohlen2002}
{Pohlen}, M., {Dettmar}, R.-J., {L{\"u}tticke}, R., \& {Aronica}, G. 2002,
  \aap, 392, 807

\bibitem[{{Pohlen} \& {Trujillo}(2006)}]{pohlen2006}
{Pohlen}, M. \& {Trujillo}, I. 2006, \aap, 454, 759

\bibitem[{{Pohlen} {et~al.}(2007){Pohlen}, {Zaroubi}, {Peletier}, \&
  {Dettmar}}]{pohlen2007}
{Pohlen}, M., {Zaroubi}, S., {Peletier}, R.~F., \& {Dettmar}, R.~. 2007,
  \mnras, 378, 594

\bibitem[{{Quinn} \& {Goodman}(1986)}]{quinn1986}
{Quinn}, P.~J. \& {Goodman}, J. 1986, \apj, 309, 472

\bibitem[{{Quinn} {et~al.}(1993){Quinn}, {Hernquist}, \&
  {Fullagar}}]{quinn1993}
{Quinn}, P.~J., {Hernquist}, L., \& {Fullagar}, D.~P. 1993, \apj, 403, 74

\bibitem[{{Roberts} \& {Haynes}(1994)}]{roberts1994}
{Roberts}, M.~S. \& {Haynes}, M.~P. 1994, \araa, 32, 115

\bibitem[{{Robertson} {et~al.}(2006{\natexlab{a}}){Robertson}, {Cox},
  {Hernquist}, {Franx}, {Hopkins}, {Martini}, \& {Springel}}]{robertson2006b}
{Robertson}, B., {Cox}, T.~J., {Hernquist}, L., {Franx}, M., {Hopkins}, P.~F.,
  {Martini}, P., \& {Springel}, V. 2006{\natexlab{a}}, \apj, 641, 21

\bibitem[{{Robertson} {et~al.}(2006{\natexlab{b}}){Robertson}, {Hernquist},
  {Cox}, {Di Matteo}, {Hopkins}, {Martini}, \& {Springel}}]{robertson2006a}
{Robertson}, B., {Hernquist}, L., {Cox}, T.~J., {Di Matteo}, T., {Hopkins},
  P.~F., {Martini}, P., \& {Springel}, V. 2006{\natexlab{b}}, \apj, 641, 90

\bibitem[{{Schaye}(2004)}]{schaye2004}
{Schaye}, J. 2004, \apj, 609, 667

\bibitem[{{Schmidt}(1959)}]{schmidt1959}
{Schmidt}, M. 1959, \apj, 129, 243

\bibitem[{{Schombert} {et~al.}(2001){Schombert}, {McGaugh}, \&
  {Eder}}]{schombert2001}
{Schombert}, J.~M., {McGaugh}, S.~S., \& {Eder}, J.~A. 2001, \aj, 121, 2420

\bibitem[{{Sellwood} {et~al.}(1998){Sellwood}, {Nelson}, \&
  {Tremaine}}]{sellwood1998}
{Sellwood}, J.~A., {Nelson}, R.~W., \& {Tremaine}, S. 1998, \apj, 506, 590

\bibitem[{{Silk} \& {Wyse}(1993)}]{silk1993}
{Silk}, J. \& {Wyse}, R.~F.~G. 1993, \physrep, 231, 293

\bibitem[{{Somerville} \& {Kolatt}(1999)}]{somerville1999}
{Somerville}, R.~S. \& {Kolatt}, T.~S. 1999, \mnras, 305, 1

\bibitem[{{Somerville} {et~al.}(2000){Somerville}, {Lemson}, {Kolatt}, \&
  {Dekel}}]{somerville2000}
{Somerville}, R.~S., {Lemson}, G., {Kolatt}, T.~S., \& {Dekel}, A. 2000,
  \mnras, 316, 479

\bibitem[{{Spergel} {et~al.}(2003)}]{spergel2003}
{Spergel}, D.~N. {et~al.} 2003, \apjs, 148, 175

\bibitem[{{Springel}(2005)}]{springel2005}
{Springel}, V. 2005, \mnras, 364, 1105

\bibitem[{{Springel} {et~al.}(2005{\natexlab{a}}){Springel}, {Di Matteo}, \&
  {Hernquist}}]{springeldimatteo2005b}
{Springel}, V., {Di Matteo}, T., \& {Hernquist}, L. 2005{\natexlab{a}}, \apjl,
  620, L79

\bibitem[{{Springel} {et~al.}(2005{\natexlab{b}}){Springel}, {Di Matteo}, \&
  {Hernquist}}]{springeldimatteo2005a}
---. 2005{\natexlab{b}}, \mnras, 361, 776

\bibitem[{{Springel} \& {Hernquist}(2002)}]{springelhernquist2002}
{Springel}, V. \& {Hernquist}, L. 2002, \mnras, 333, 649

\bibitem[{{Springel} \& {Hernquist}(2003)}]{springelhernquist2003}
---. 2003, \mnras, 339, 289

\bibitem[{{Thilker} {et~al.}(2005)}]{thilker2005}
{Thilker}, D.~A. {et~al.} 2005, \apjl, 619, L79

\bibitem[{{Toomre}(1977)}]{toomre1977}
{Toomre}, A. 1977, in Evolution of Galaxies and Stellar Populations, ed. B.~M.
  {Tinsley} \& R.~B. {Larson}, 401

\bibitem[{{Toomre} \& {Toomre}(1972)}]{toomre1972}
{Toomre}, A. \& {Toomre}, J. 1972, \apj, 178, 623

\bibitem[{{Trujillo} \& {Pohlen}(2005)}]{trujillo2005}
{Trujillo}, I. \& {Pohlen}, M. 2005, \apjl, 630, L17

\bibitem[{{van der Kruit}(1979)}]{vanderKruit1979}
{van der Kruit}, P.~C. 1979, \aaps, 38, 15

\bibitem[{{van der Kruit}(2001)}]{vanderKruit2001}
{van der Kruit}, P.~C. 2001, in ASP Conf. Ser. 230: Galaxy Disks and Disk
  Galaxies, ed. J.~G. {Funes} \& E.~M. {Corsini}, 119--126

\bibitem[{{Velazquez} \& {White}(1999)}]{velazquez1999}
{Velazquez}, H. \& {White}, S.~D.~M. 1999, \mnras, 304, 254

\bibitem[{{Walker} {et~al.}(1996){Walker}, {Mihos}, \&
  {Hernquist}}]{walker1996}
{Walker}, I.~R., {Mihos}, J.~C., \& {Hernquist}, L. 1996, \apj, 460, 121

\bibitem[{{Wyse} {et~al.}(2006)}]{wyse2006}
{Wyse}, R.~F.~G. {et~al.} 2006, \apjl, 639, L13

\end{thebibliography}

\end{document}